\documentclass[apjl]{emulateapj}
\usepackage{psfig,amsfonts,amsmath,graphicx,natbib,apjfonts}
\def\lsr{LSR\,1835+32}

\def\tvlm{TVLM\,513-46546}
\def\2m0746{2M\,0746+20}
\def\swift{{\it Swift}}

\def\har{1}
\def\mcgill{2}
\def\sin{3}
\def\ucb{4}
\def\noao{5}
\def\udel{6}
\def\steward{7}
\def\iac{8}
\def\ucf{9}

\begin{document}

\title{Periodic Radio and H$\alpha$ Emission from the L Dwarf Binary
2MASSW J0746425+200032: Exploring the Magnetic Field Topology and
Radius of an L Dwarf}

\author{
E.~Berger\altaffilmark{\har},
R.~E.~Rutledge\altaffilmark{\mcgill},
N.~Phan-Bao\altaffilmark{\ucf},
G.~Basri\altaffilmark{\ucb},
M.~S.~Giampapa\altaffilmark{\noao},
J.~E.~Gizis\altaffilmark{\udel},
J.~Liebert\altaffilmark{\steward},
E.~Mart{\'{\i}}n\altaffilmark{\iac,}\altaffilmark{\ucf},
and T.~A.~Fleming\altaffilmark{\steward},
}

\altaffiltext{\har}{Harvard-Smithsonian Center for Astrophysics, 60
Garden Street, Cambridge, MA 02138}


\altaffiltext{\sin}{Institute of Astronomy and Astrophysics, Academia
Sinica, P.O.~Box 23-141, Taipei 10617, Taiwan, R.O.C.}

\altaffiltext{\ucb}{Astronomy Department, University of California,
Berkeley, CA 94720}

\altaffiltext{\noao}{National Solar Observatory, National Optical
Astronomy Observatories, Tucson, AZ 85726}

\altaffiltext{\udel}{Department of Physics and Astronomy, University
of Delaware, Newark, DE 19716}

\altaffiltext{\steward}{Department of Astronomy and Steward
Observatory, University of Arizona, 933 North Cherry Avenue, Tucson,
AZ 85721}

\altaffiltext{\iac}{Instituto de Astrof{\'{\i}}sica de Canarias, C/
V{\'{\i}}a L\'actea s/n, E-38200 La Laguna, Tenerife, Spain}

\altaffiltext{\ucf}{University of Central Florida, Department of
Physics, PO Box 162385, Orlando, FL 32816}

\begin{abstract} We present an 8.5-hour simultaneous radio, X-ray, UV,
and optical observation of the L dwarf binary 2MASSW J0746425+200032.
We detect strong radio emission, dominated by short-duration periodic
pulses at 4.86 GHz with $P=124.32\pm 0.11$ min.  The stability of the
pulse profiles and arrival times demonstrates that they are due to the
rotational modulation of a $B\approx 1.7$ kG magnetic field.  A
quiescent non-variable component is also detected, likely due to
emission from a uniform large-scale field.  The H$\alpha$ emission
exhibits identical periodicity, but unlike the radio pulses it varies
sinusoidally and is offset by exactly 1/4 of a phase.  The sinusoidal
variations require chromospheric emission from a large-scale field
structure, with the radio pulses likely emanating from the magnetic
poles.  While both light curves can be explained by a rotating
mis-aligned magnetic field, the 1/4 phase lag rules out a symmetric
dipole topology since it would result in a phase lag of 1/2 (poloidal
field) or zero (toroidal field).  We therefore conclude that either
(i) the field is dominated by a quadrupole configuration, which can
naturally explain the 1/4 phase lag; or (ii) the H$\alpha$ and/or
radio emission regions are not trivially aligned with the field.
Regardless of the field topology, we use the measured period along
with the known rotation velocity ($v{\rm sin}i\approx 27$ km
s$^{-1}$), and the binary orbital inclination ($i\approx 142^\circ$),
to derive a radius for the primary star of $0.078\pm 0.010$ R$_\odot$.
This is the first measurement of the radius of an L dwarf, and along
with a mass of $0.085\pm 0.010$ M$_\odot$ it provides a constraint on
the mass-radius relation below 0.1 M$_\odot$.  We find that the radius
is about $30\%$ smaller than expected from theoretical models, even
for an age of a few Gyr.  The origin of this discrepancy is either a
breakdown of the models at the bottom of the main sequence, or a
significant mis-alignment between the rotational and orbital axes.
\end{abstract}
 
\keywords{radio continuum:stars --- stars:activity --- stars:low-mass,
brown dwarfs --- stars:magnetic fields}

\section{Introduction}
\label{sec:intro}

Radio observations conducted over the past several years have
uncovered a substantial fraction of magnetically active low mass stars
and brown dwarfs.  Both quiescent and flaring emission are present,
with luminosities that remain unchanged down to at least spectral type
$\sim {\rm L3}$, in contrast to the declining activity seen in X-rays
and H$\alpha$
\citep{bbb+01,ber02,brr+05,bp05,ber06,ohb+06,had+06,adh+07,pol+07,hbl+07,aob+07,adh+08,bgg+08,bbg+08,had+08}.
Depending on the nature of the radio emission mechanism, the inferred
magnetic field strengths are $\sim 0.1-3$ kG with order unity filling
factors \citep{ber06,had+08}.  Long term monitoring of several objects
in the spectral type range M8--L3.5 has further shown that the fields
are generally stable for at least several years
(\citealt{brr+05,ber06,bgg+08}; but see also \citealt{adh+07}),
providing an important constraint on the lifetime of magnetic dynamos
in fully convective objects.  The strength, scale, and stability of
the inferred fields is in good agreement with recent results from
phase-resolved spectropolarimetry (spectral type $\sim {\rm M4}$;
\citealt{dfc+06,mdp+08}) and Zeeman broadening of FeH lines (spectral
type $<{\rm M9}$; \citealt{rb06,rb07}), as well as with the most
recent numerical dynamo simulations (e.g., \citealt{bro08}).

Equally important, three ultracool dwarfs to date have been observed
to produce periodic radio emission, with periods of 184 min (2MASS
J00361617+1821104; \citealt{brr+05}), 118 min
(\tvlm\footnotemark\footnotetext{This object also exhibits periodic
H$\alpha$ emission, with the same period as observed in the radio
\citep{bgg+08}.}; \citealt{hbl+07}), and 170 min (LSR\,1835+3259;
\citealt{had+08}).  In the case of 2M\,0036+18 the periodic emission
is sinusoidal, while in the latter two objects it is in the form of
short duration pulses (duty cycle of a few percent).  In all three
cases the observed periods are in good agreement with the known
rotation velocities ($v{\rm sin}i$), indicating that the radio
periodicity traces the stellar rotation.

As a result, in addition to allowing a measurement of the magnetic
field properties, radio observations provide a unique opportunity to
measure the radii of late-M and L dwarfs through the combination of
rotation period and velocity.  However, since only $v{\rm sin}i$
values are known for these objects, there is an inherent degeneracy
between the radius and the inclination of the rotation axis, $i$.
Using a typical range of $R\approx 0.09-0.11$ R$_\odot$ for late-M and
L dwarfs, the allowed range of inclinations span a relatively wide
range of $\sim 60-90^\circ$ \citep{brr+05,had+08}.  To break this
degeneracy, and thus measure the radius directly, it is desirable to
observe objects for which the inclination can be estimated.  This is
the case for binary systems with well determined orbital parameters if
we make the reasonable assumption that the rotation and orbital axes
are aligned \citep{hal94}.  Along with an estimate of the mass, we can
thus place constraints on the mass-radius relation for ultracool
dwarfs.

Here we present simultaneous radio, X-ray, optical, and UV
observations of one such system, the L dwarf binary 2MASSW
J0746425+200032 (hereafter, \2m0746).  These observations are part of
a long-term project to study the field properties of ultracool dwarfs
\citep{brr+05,bgg+08,bbg+08}.  While no X-ray or UV emission are
detected, the radio and H$\alpha$ emission exhibit clear periodicity,
which along with the known orbital inclination and $v{\rm sin}i$ allow
us to explore for the first time the radius and magnetic field
topology of an L dwarf.

\section{Observations}
\label{sec:obs}

\subsection{Target Selection and Properties}

We targeted the L dwarf binary \2m0746\ (L0+L1.5) due to the
availability of its orbital parameters and rotation velocity, as well
as previous detections in H$\alpha$.  The binary is located at a
distance of only $12.2$ pc \citep{dhv+02}.  The orbital inclination
and total binary mass are $142\pm 3^\circ$ and
$0.146^{+0.016}_{-0.006}$ M$_\odot$, respectively \citep{bdk+04}.  The
rotation velocities quoted in the literature are $26\pm 3$ km s$^{-1}$
\citep{bai04}, 24 km s$^{-1}$ \citep{rkl+02}, 31 km s$^{-1}$
\citep{rb08}, and 28 km s$^{-1}$ (C.~Blake priv.~comm.)  We therefore
adopt an average value of $v{\rm sin}i=27\pm 3$ km s$^{-1}$.  The
bolometric luminosity of \2m0746\ is $L_{\rm bol}=10^{-3.64\pm 0.06}$
L$_\odot$ \citep{vhl+04}.  We note that there is still no agreement as
to whether the secondary member of the binary is a low mass star or a
brown dwarf \citep{bdk+04,gr06}.

Previous H$\alpha$ detections revealed an equivalent width range of
$\approx 1.2-2.4$ \AA, or ${\rm log}\,(L_{\rm H\alpha}/L_{\rm
bol})\approx -5.3$ to $-5.2$ \citep{rkg+00,rkl+02,scb+07,rb08}.  No
radio emission was previously detected at 8.46 GHz to a $3\sigma$
limit of $\lesssim 48$ $\mu$Jy \citep{ber06}, but subsequent to the
observations presented here, \citet{adh+08} published a radio
detection at 4.86 GHz with $F_\nu=286\pm 24$ $\mu$Jy, based on a
2-hour observation.  They further detect the possible emergence of a
flare in the last few minutes of their observation.

The simultaneous observations presented here were conducted on 2008
February 22 UT for a total of 8.4 hr in the radio (02:11--10:34 UT),
8.83 hr in the X-rays (02:20--11:10 UT), and 7.4 hr in the optical
(05:43--13:09 UT).  Observations with the \swift\ UV/Optical Telescope
(UVOT) took place intermittently between 01:32 and 09:54 UT with a
total on-source exposure time of 2.15 hr.

\subsection{Radio} 
\label{sec:rad}

Very Large Array\footnotemark\footnotetext{The VLA is operated by the
National Radio Astronomy Observatory, a facility of the National
Science Foundation operated under cooperative agreement by Associated
Universities, Inc.} observations were obtained simultaneously at 4.86
and 8.46 GHz in the standard continuum mode with $2\times 50$ MHz
contiguous bands.  Thirteen antennas were used at each frequency in
the BnC array configuration.  Scans of 295 s on source were
interleaved with 55 s scans on the phase calibrator J0738+177.  The
flux density scale was determined using the extragalactic source
3C\,286 (J1331+305).  Data reduction and analysis follow the
procedures outlined in \citet{bgg+08} and \citet{bbg+08}.  We detect a
source coincident with the position\footnotemark\footnotetext{We take
into account the known proper motion of 370 mas yr$^{-1}$ at a
position angle of $264^\circ$ \citep{scb+07}.} of \2m0746\ at both
frequencies.

\subsection{Optical Spectroscopy} 
\label{sec:optical}

We used\footnotemark\footnotetext{Observations were obtained as part
of program GN-2008A-Q-11.} the Gemini Multi-Object Spectrograph (GMOS;
\citealt{hja+04}) mounted on the Gemini-North 8-m telescope with the
B600 grating at a central wavelength of 5250 \AA, and with a $1''$
slit.  The individual 300-s exposures were reduced using the {\tt
gemini} package in IRAF (bias subtraction, flat-fielding, and sky
subtraction), and the wavelength solution was determined from CuAr arc
lamps.  The spectra cover $3840-6680$ \AA\ at a resolution of about 5
\AA.  A total of 77 exposures were obtained, with an on-source
efficiency of $94\%$.  We detect H$\alpha$ emission in all the
individual spectra.

\subsection{X-Rays}
\label{sec:xrays}

Observations were performed with the Chandra/ACIS-S3
backside-illuminated chip for a total of 29.46 ks.  The data were
analyzed using CIAO version 3.4, and counts were extracted in a $2''$
radius circle centered on the position of \2m0746.  We find only 2
counts, with 1.5 counts expected from the background as determined
from annuli centered on the source position.  Thus, the resulting
upper limit is about 7 counts ($95\%$ confidence level).  Using an
energy conversion factor of $1\,{\rm cps}=3.8\times 10^{-12}$ erg
cm$^{-2}$ s$^{-1}$ (appropriate for a 1 keV Raymond-Smith plasma model
in the $0.2-2$ keV range) we find $F_X<9.0\times 10^{-16}$ erg
cm$^{-2}$ s$^{-1}$, or $L_X/L_{\rm bol}\lesssim 10^{-4.7}$.

\subsection{Ultraviolet}

Data were obtained with the \swift/UVOT in the UVW1 filter
($\lambda_{\rm eff}\approx 2600$ \AA), as a series of 6 images with
exposure times ranging from 460 to 1625 s.  No source is detected at
the position of \2m0746\ in any of the individual exposures or in the
combined 2.15 hr image.  Photometry of the combined image results in a
limit of $F_\lambda<2.8\times 10^{-18}$ erg cm$^{-2}$ s$^{-1}$
\AA$^{-1}$ in a $2''$ aperture.  The ratio of UV to bolometric
luminosity is $\lambda L_\lambda/L_{\rm bol}\lesssim 10^{-3.8}$.

\section{Multi-Wavelength Periodic Emission} 
\label{sec:prop}

\subsection{Radio Emission}

The radio emission from \2m0746\ has an average flux density of
$304\pm 15$ $\mu$Jy at 4.86 GHz, and $154\pm 14$ $\mu$Jy at 8.46 GHz.
This is the first detection of the object at 8.46 GHz, with an
increase by at least a factor of three compared to previous limits
from June 2002 \citep{ber06}.  The average fraction of circular
polarization is $\lesssim 15\%$ ($3\sigma$) and $35\pm 10\%$,
respectively.  The resulting average luminosities are $L_\nu
(4.86)=(5.4\pm 0.3)\times 10^{13}$ and $L_\nu (8.46)=(2.8\pm
0.3)\times 10^{13}$ erg cm$^{-2}$ s$^{-1}$ Hz$^{-1}$, and the ratio
relative to the bolometric luminosity is $\nu L_\nu/L_{\rm bol}\approx
10^{-6.55}$.  These values are similar to those of previously-detected
late-M and L dwarfs \citep{ber06}.

\begin{figure}
\centerline{\psfig{file=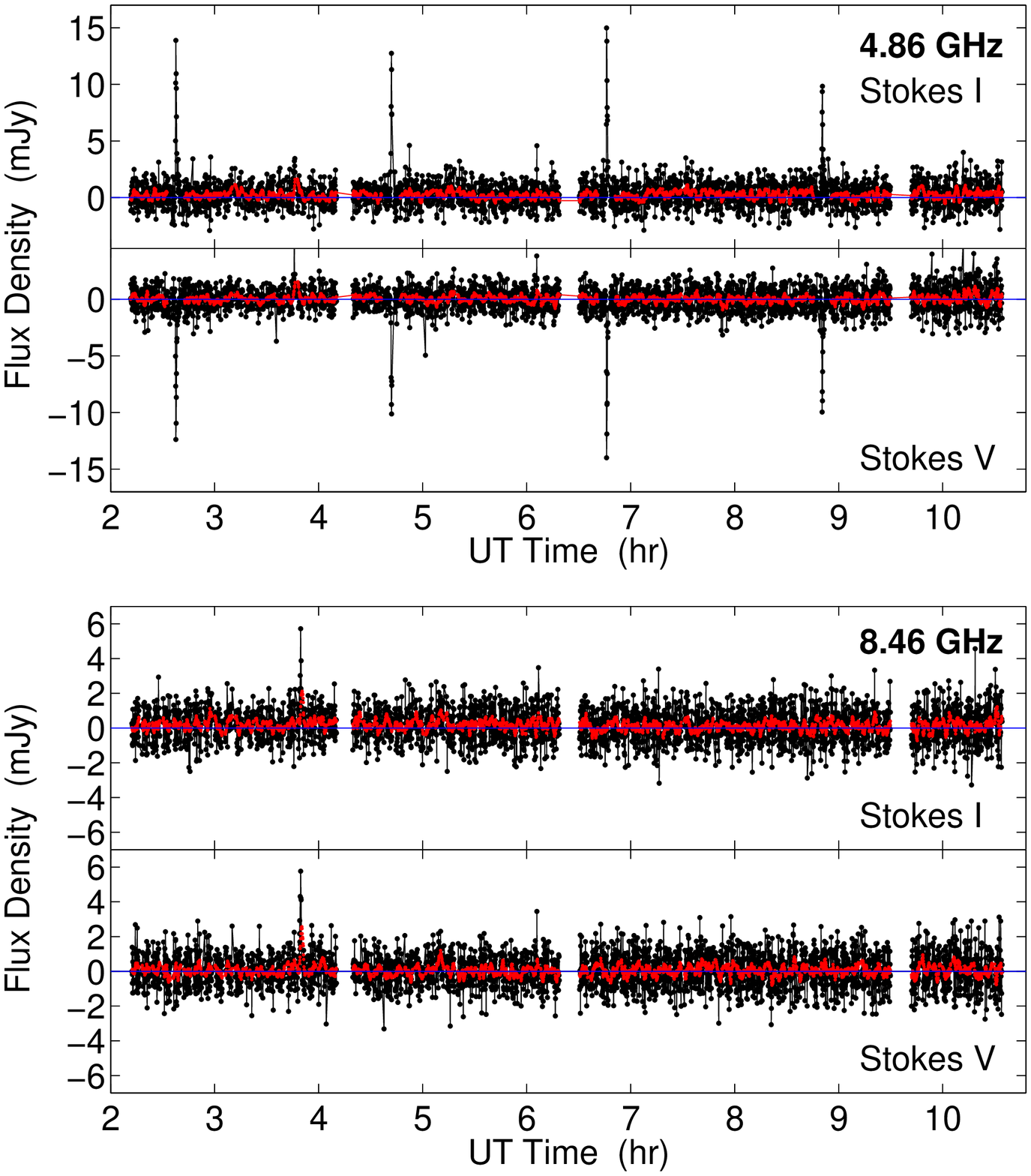,width=6.0in}} 
\caption{Radio light curves of \2m0746\ at 4.86 GHz (top) and 8.46 GHz
(bottom) in stokes I (total intensity) and stokes V (circular
polarization; negative values indicate left-handed polarization).  The
data are shown at the native 5-s time resolution (black points) and
smoothed with a 1-min boxcar (red points).  The radio emission at 4.86
GHz is dominated by bright, $\sim 100\%$ circularly polarized, and
periodic bursts ($P=124.32\pm 0.11$ min), with no corresponding
emission at 8.46 GHz.  These properties point to coherent emission.
In addition, we detect quiescent, non-variable emission at both
frequencies, with a spectral index indicative of gyrosynchrotron
radiation.
\label{fig:radio}}
\end{figure}

The radio light curves are shown in Figure~\ref{fig:radio}.  The most
striking aspect of the radio emission is a set of bright ($10-15$
mJy), short duration (1.2 min), circularly polarized ($\sim 100\%$),
and {\it periodic} pulses at 4.86 GHz.  The period measured from the
well-defined peaks\footnotemark\footnotetext{The UT times of the four
peaks are 02:37:37.6, 04:41:57.5, 06:46:07.3, and 08:50:37.7.}  of the
four detected pulses is $124.32\pm 0.11$ min.  No corresponding
emission is detected at 8.46 GHz, indicating that the fractional
bandwidth is $\delta\nu/\nu\lesssim 0.7$.  The pulses are detected at
both intermediate frequencies of the 4.86 GHz band (4885 MHz and 4835
MHz), with no clear time delay implying that $\delta\nu/\nu\gtrsim
0.02$.

\begin{figure}
\centerline{\psfig{file=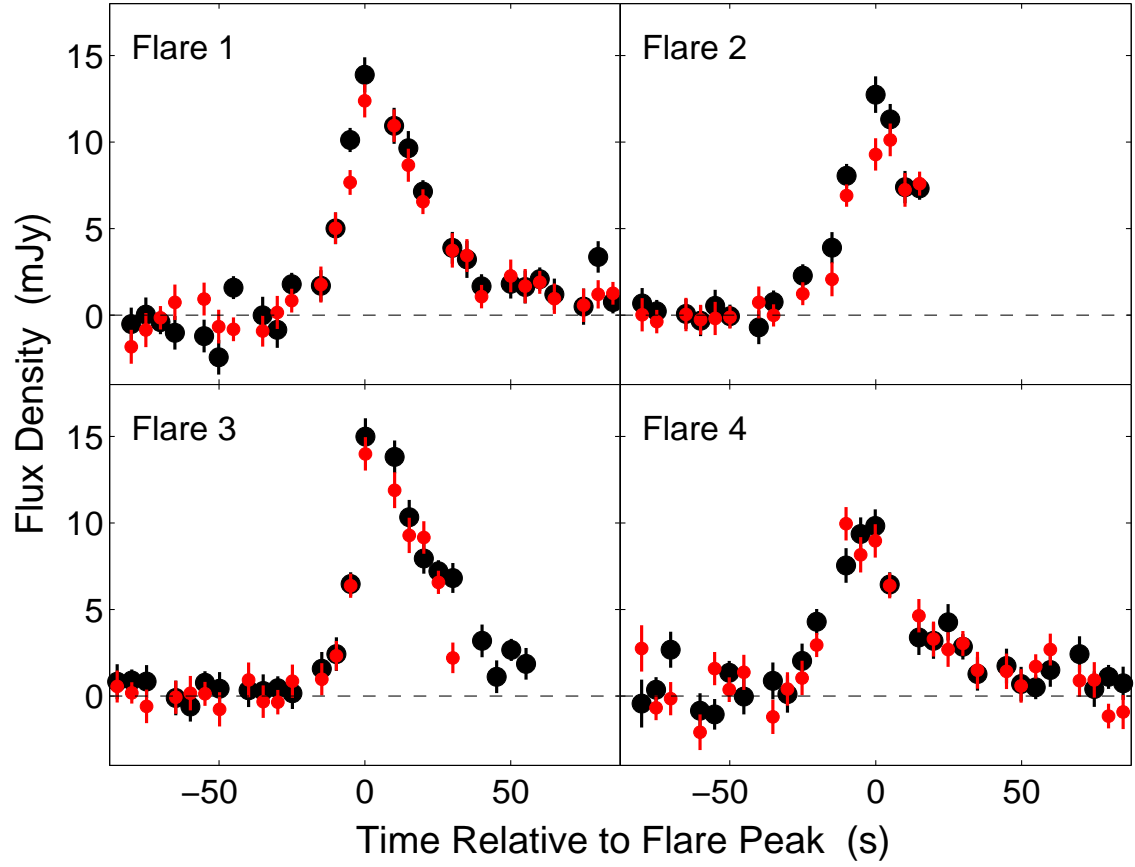,width=6.0in}}
\caption{Zoom-in view on the four bursts detected at 4.86 GHz.  Time
is measured relative to the peak of each burst.  Both total intensity
(black) and circular polarization (red) are shown, with the latter
inverted for ease of comparison.  The bursts are nearly $100\%$
circularly polarized.  The rise time of the bursts is about 20 s,
while the decay time is about 40 s.
\label{fig:flares}}
\end{figure}

The detailed temporal profiles of the pulses in total intensity and
circular polarization are shown in Figure~\ref{fig:flares}.  The
profiles are similar in all four cases, with a rise time of about 20
s, followed by a decline timescale of 40 s.  The high degree of
stability of the period and emission properties demonstrate that the
pulsing activity is the result of stellar rotation, rather than
genuine episodic flares.  In this context, the duty cycle of only
$0.8\%$ implies that the radio emitting region has an azimuthal scale
of only $\sim 0.1$ R$_*$ (assuming a height above the surface of about
$0.5-1$ R$_*$; \citealt{had+08}).  Furthermore, since the measured
period roughly agrees with the rotation velocity of \2m0746\
(\S\ref{sec:obs}), we conclude that only one pulse is observed per
rotation.  Along with the uniform sense of circular polarization, this
indicates that we observe only one of the magnetic poles.

The properties of the periodic pulses are similar to those observed in
\tvlm\ and \lsr\ \citep{hbl+07,had+08}, and point to a coherent
emission process, most likely the electron cyclotron maser (e.g.,
\citealt{tre06}).  In this framework, the radiation is produced
primarily at the electron cyclotron frequency, $\nu_c\approx 2.8\times
10^6\,B$ Hz, indicating that the magnetic field strength of \2m0746\
is $B\approx 1.7$ kG.  We can also place a limit on the electron
density using the constraint that the plasma frequency, $\nu_p\approx
9\times 10^{3}\,n_e^{1/2}$ Hz, has to be lower than $\nu_c$.  We thus
find, $n_e\lesssim 3\times 10^{11}$ cm$^{-3}$.

A single short duration burst is also detected at 8.46 GHz (03:49:36
UT; Figure~\ref{fig:radio}), with a peak flux of about 6 mJy, an
identical duration to the 4.86 GHz pulses, and $\sim 100\%$ right
circular polarization.  No corresponding emission is detected at 4.86
GHz.  In the context of electron cyclotron maser emission, this burst
requires a stronger magnetic field component of $B\approx 3$ kG.  The
burst is delayed by 72 min (or $0.58\times P$) relative to the
preceding 4.86 GHz pulse, indicating that its emission region is not
trivially related to that of the periodic pulses.  This, and the lack
of periodicity, suggest that the 8.46 GHz burst is either the result
of transitory field dissipation (a genuine flare), or that the
physical conditions in its emission region (field strength and/or
density) vary rapidly with time so as to suppress the emerging 8.46
GHz radiation during subsequent rotations.  In either case, it is
clear that the magnetic field strength ranges by at least a factor of
two over the stellar surface and/or corona.

Eliminating the contribution of the short duration pulses, we find
baseline quiescent emission of $224\pm 15$ $\mu$Jy (4.86 GHz) and
$149\pm 15$ $\mu$Jy (8.46 GHz) with no clear sign of variability
(Figure~\ref{fig:radio}).  The quiescent emission is distinct from the
narrow-band pulses since it is detected at both 4.86 and 8.46 GHz and
it does not vary with the stellar rotation.  The wide frequency range
($\delta\nu/\nu\sim 1$) and inferred spectral index of $-0.7\pm 0.3$
are typical of gyrosynchrotron emission from a power law distribution
of electrons with an energy index of $p\approx 2.4$ \citep{gud02}.
The lack of variability during 8.4 hours (or four rotations) suggests
that the quiescent emission arises from particle acceleration in a
uniform large-scale field component with order unity covering
fraction.

\subsection{H$\alpha$ Emission}
\label{sec:halpha}

\begin{figure}
\centerline{\psfig{file=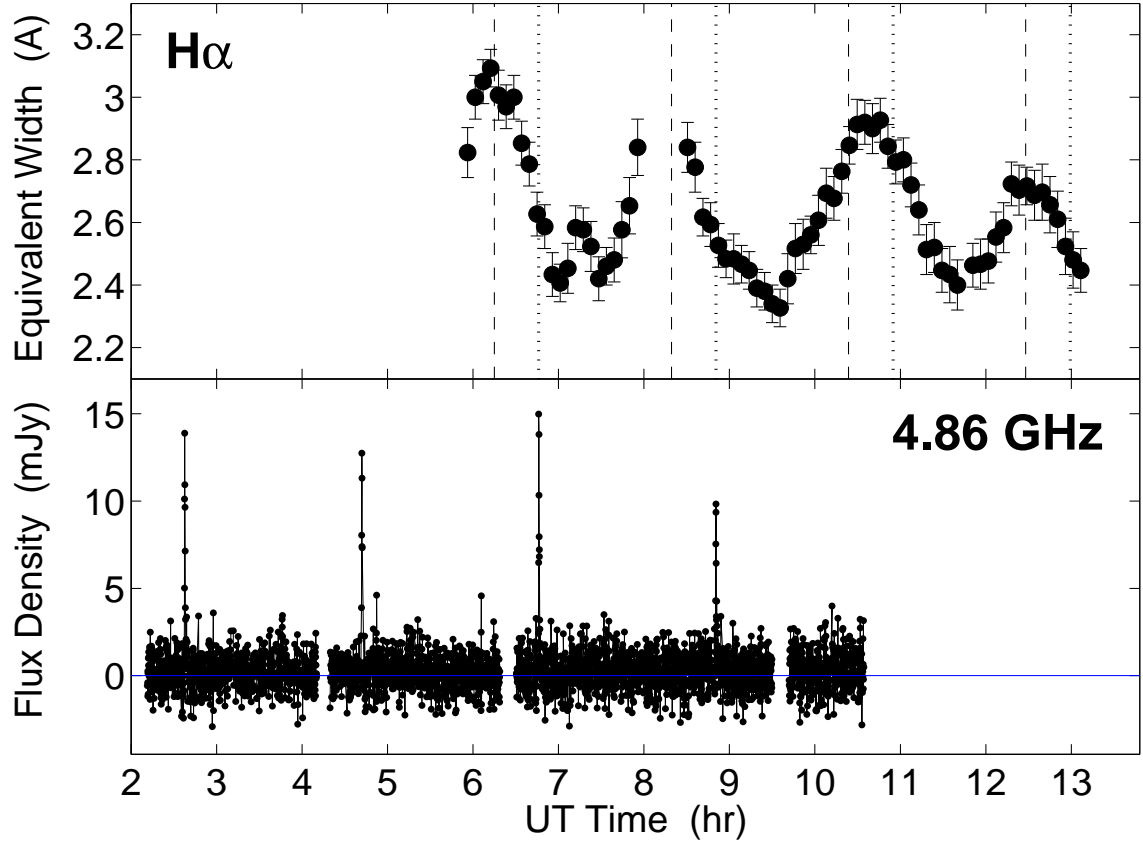,width=6.0in}}
\caption{H$\alpha$ equivalent width light curve (top) in comparison to
the 4.86 GHz total intensity light curve (bottom).  The H$\alpha$
emission is clearly sinusoidal, with the same period as that measured
from the radio bursts.  The peaks of the H$\alpha$ light curve (dashed
lines) lead the radio bursts (dotted lines) by exactly 1/4 of a
period.  These properties, along with the non-zero minima, point to
emission from a large-scale chromospheric structure whose projected
solid angle varies with the rotation of \2m0746.
\label{fig:halpha}} 
\end{figure}

The H$\alpha$ light curve is shown in Figure~\ref{fig:halpha}.  The
equivalent width ranges from 2.4 to 3.1 \AA, with a mean value of
about 2.7 \AA.  More importantly, the light curve is clearly periodic,
with $P=126\pm 10$ min in excellent agreement with the radio
periodicity.  This indicates that the radio and H$\alpha$ emission
arise from the same binary member.  However, despite the identical
periods, two clear differences are present between the H$\alpha$ and
radio light curves, which provide additional constraints on the
geometry of the magnetic field.

First, the H$\alpha$ light curve is sinusoidal as opposed to the $\sim
1\%$ duty cycle of the radio pulses.  This indicates that the
H$\alpha$ emission is produced by chromospheric plasma that covers a
substantial fraction (though less than 100\%) of the stellar surface
and rotates relative to our line of sight.  A small enhancement is
ruled out since it would result in a sharp rise and decline through
ingress and egress, respectively, and a flat-topped peak.  In
addition, the non-zero minimum equivalent width ($\approx 2.4$ \AA)
indicates that the sinusoidal variations are most likely due to a
combination of rotation and field orientation effects, such that at
any given time some fraction of the H$\alpha$-emitting chromosphere is
visible, with the maximum projected solid angle corresponding to the
light curve peaks (see \S\ref{sec:geom}).

Second, while the radio and H$\alpha$ periods are identical, the phase
of the two light curves differs (Figure~\ref{fig:halpha}).  In
particular, we find that the H$\alpha$ peaks lead the radio pulses by
about 31 min, corresponding to 1/4 of the period (or, equivalently 1/4
of a rotation).  This lag indicates that the narrow beam which gives
rise to the radio pulses is offset by about $90^\circ$ relative to the
central axis of the chromospheric geometry.  As we show in the
following section, such a $90^\circ$ offset cannot be accommodated in a
simple dipole magnetic field model, unless there is a significant
breaking of the symmetry between the field and emission regions.

We have previously observed similar periodic H$\alpha$ emission from
the M8.5 dwarf \tvlm, with $P\approx 2$ hr well-matched to the
rotation velocity of $v{\rm sin}i\approx 60$ km s$^{-1}$
\citep{bgg+08}.  This object has also been shown to produce radio
bursts on a separate occasion \citep{hbl+07}.  In the case of \lsr,
on the other hand, despite the presence of periodic radio bursts
\citep{had+08} no periodic H$\alpha$ emission was evident in a
separate 5.4 hr observation \citep{bbg+08}.  It is thus possible that
the magnetic field topology and stability timescale, as well as the
viewing and magnetic axis orientation, play a role in determining the
correlation (or lack thereof) between the periodic radio and H$\alpha$
signals.

\section{The Magnetic Field Geometry}
\label{sec:geom}

Taking into account the properties of the radio and H$\alpha$
emission, we now investigate the magnetic field topology.  This
topology has to satisfy the following requirements: (i) the H$\alpha$
emission is due to a large scale feature; (ii) the non-zero minimum of
the H$\alpha$ light curve requires some fraction of the active
chromosphere to be visible at all times; (iii) the short duration
radio pulses arise from a narrow region; and (iv) the radio pulses lag
the H$\alpha$ peaks by 1/4 of the period.

\begin{figure}
\centerline{\psfig{file=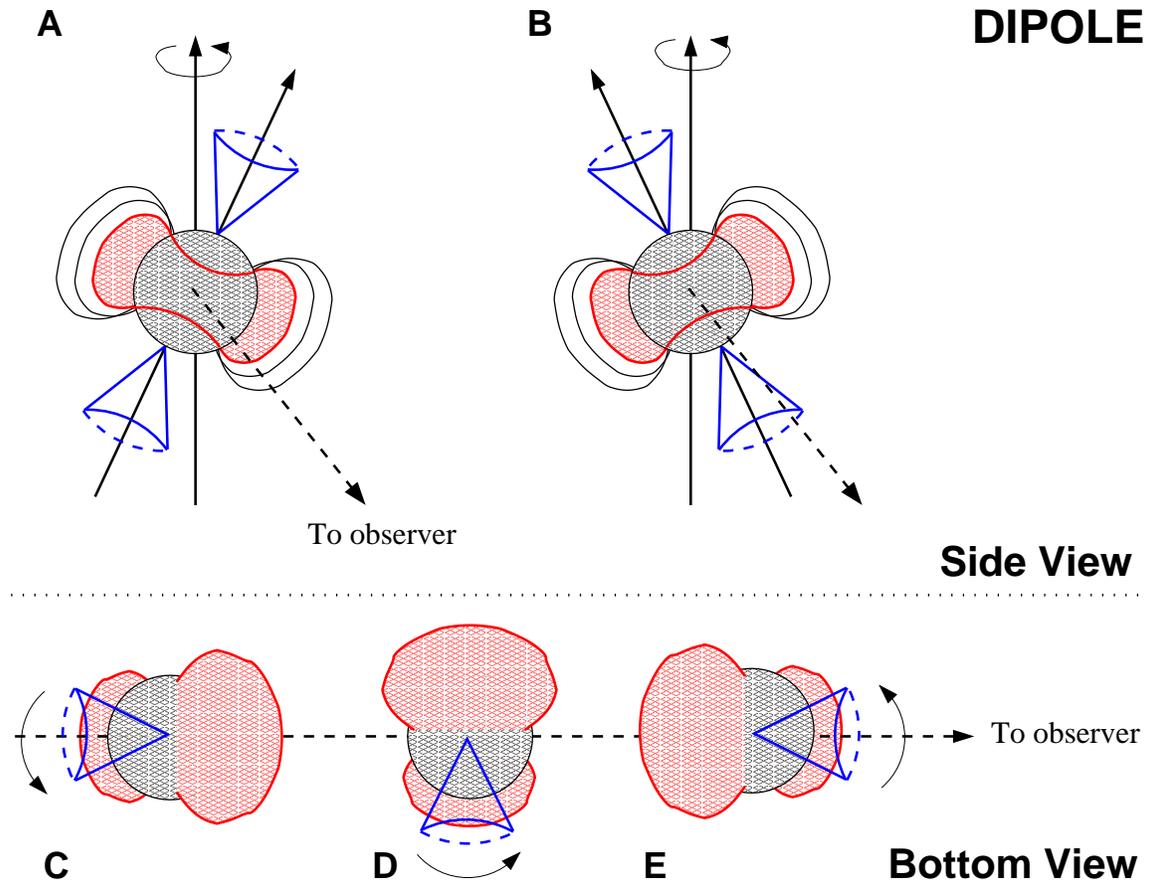,width=6.0in}}
\caption{A dipole poloidal field topology for \2m0746, and its
orientation relative to the line of sight.  This configuration leads
to radio bursts (blue) and sinusoidal H$\alpha$ emission (red).
However, as shown in both sets of projections, we expect a 1/2 phase
lag between the two bands, with the minimum projected solid angle of
the field (i.e., minimum H$\alpha$ emission) coinciding with the radio
bursts.  In the alternative toroidal configuration the H$\alpha$ and
radio peaks will be aligned since both are produced in the same polar
region.
\label{fig:geom1}} 
\end{figure}

The first scenario we explore is a poloidal field with a simple dipole
geometry.  Such fields have been inferred for convective mid-M dwarfs,
such as V374 Peg, from phase-resolved spectropolarimetry
\citep{dfc+06,mdp+08}.  This field geometry, and our line of sight
orientation, are shown in Figure~\ref{fig:geom1}.  We assume that the
inclination of the rotation axis relative to our line of sight is
identical to the orbital axis, $i=142^\circ$.  In order to explain the
radio pulses with a small duty cycle the magnetic axis has to be
tilted relative to the rotation axis so that the beam emerging from
the radial field at the poles sweeps into out line of sight once per
rotation (a ``pulsar configuration").  This configuration also
explains the H$\alpha$ modulation since the solid angle subtended by
the poloidal field that is projected along our line of sight varies
sinusoidally as the object rotates.  Specifically, as the field is
tilted towards our line of sight we observe maximum H$\alpha$ emission
(see panels {\it A} and {\it B} of Figure~\ref{fig:geom1} for the
geometry corresponding to the maximum and minimum H$\alpha$ emission,
respectively).

Unfortunately, while this simple model explains both the radio
pulsations and the H$\alpha$ light curve, it predicts a {\it 1/2 phase
lag} between the two light curves, rather than the observed 1/4 phase
lag.  This is simply because the same phase of the rotation that
orients the magnetic pole toward our line of sight (resulting in a
radio burst), also projects the minimum solid angle of the dipole
field (i.e., minimum H$\alpha$ emission).  This effect is independent
of our choice of orbital inclination (Figure~\ref{fig:geom1}).

Alternatively, a toroidal dipole configuration, with the polar caps
producing the brightest H$\alpha$ emission, will also produce radio
bursts and sinusoidal H$\alpha$ emission.  However, in this
configuration the two light curves would be exactly in phase since
when the polar region is oriented toward our line of sight we will
observe both an enhancement of the H$\alpha$ emission and a radio
pulse.

\begin{figure}
\centerline{\psfig{file=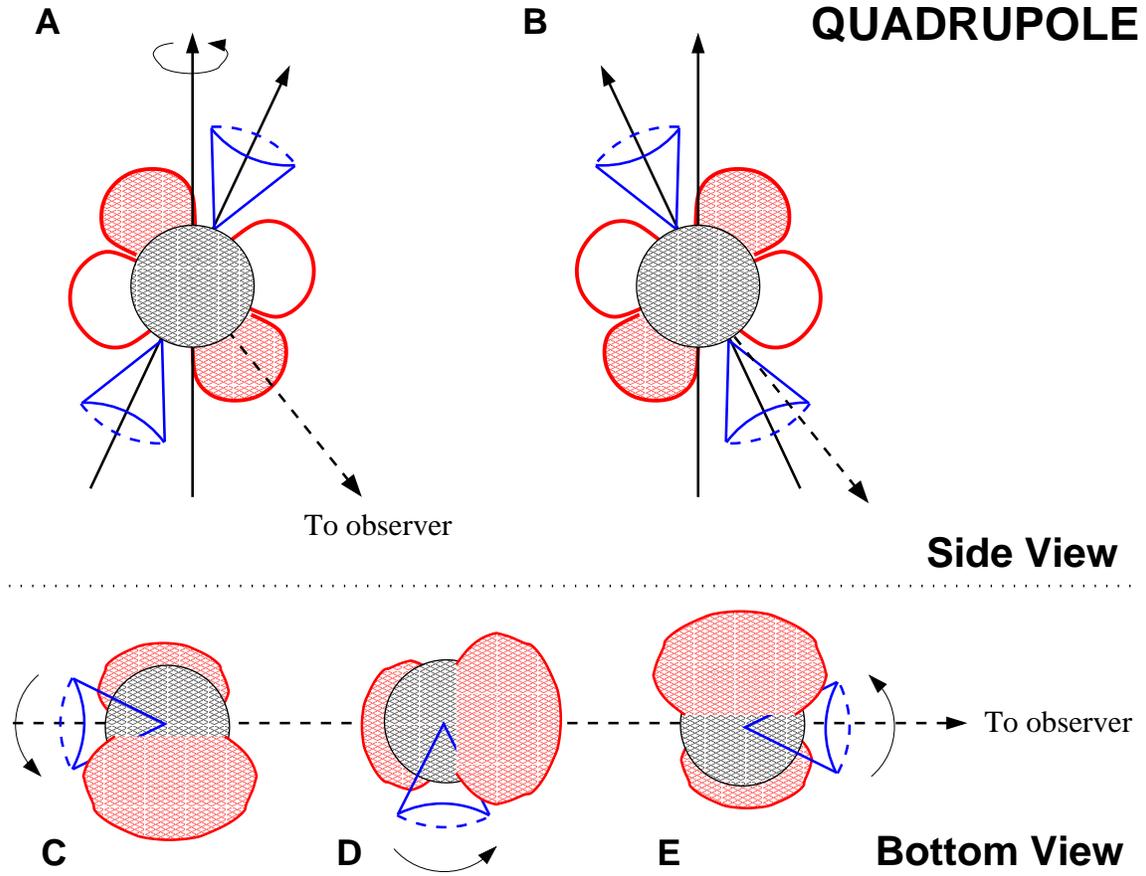,width=6.0in}}
\caption{A quadrupole field configuration for \2m0746, and its
orientation relative to the line of sight.  This configuration
reproduces both the light curve shapes and the 1/4 phase lag.
However, it requires non-uniform H$\alpha$ surface brightness, with
two of the quadrupole lobes (on opposite hemispheres) producing
stronger H$\alpha$ emission.
\label{fig:geom2}} 
\end{figure}

Having rejected the simple dipole poloidal and toroidal
configurations, we are left with two possibilities to explain the 1/4
phase lag and the radio/H$\alpha$ light curves.  First, the magnetic
field configuration is more complex and dominated by higher order
multipoles.  Or second, there is a breaking of the symmetry either in
the alignment between the poles and radio pulses, or between the
poloidal/toroidal structure and the H$\alpha$ emission.  In the former
case, the simplest possibility is that the field is quadrupolar,
leading naturally to the possibility of a 1/4 phase lag (as opposed to
0 or 1/2 phase lag for the dipole field).  However, the quadrupole
cannot produce uniform H$\alpha$ emission since this would result in a
non-varying light curve.  Instead, two of the quadrupole ``lobes" have
to produce enhanced H$\alpha$ emission -- for example, one each in the
northern and southern hemispheres.  This configuration is shown in
Figure~\ref{fig:geom2}, and it appears to explain all of the available
observations.

The alternative hypothesis of symmetry breaking can accommodate several
possibilities.  In the simplest scenario, the radio emission does not
emerge uniformly from the polar caps, but is instead produced in a
localized region at lower latitude (a ``hot spot configuration").  As
the object rotates the hot spot moves in and out of our line of sight.
This may work in both the toroidal and poloidal dipole configurations,
since both can explain the sinusoidal H$\alpha$ variability.  However,
the arbitrary location of the hot spot does not trivially explain the
exact 1/4 phase lag observed here.  

More complex possibilities exist for both the high-order multipole
scenario and the broken symmetry scenario.  However, the stability of
the radio pulses and the overall smooth H$\alpha$ light curve suggest
that the dominant topology is not significantly more complex than the
scenarios we explored here.

To summarize, we find that the 1/4 phase lag between the radio and
H$\alpha$ light curves eliminates simple and symmetric dipole field
configurations in which the radio emission is produced at the poles,
and the H$\alpha$ emission arises from plasma confined by the dipole
field; in this configuration we expect the phase lag to be either zero
(toroidal field) or 1/2 (poloidal field).  The simplest alternative is
to either: (i) retain the polar origin of the radio emission, but
appeal to a quadrupole field with non-uniform H$\alpha$ emission; or
(ii) retain the dipole configuration as the origin of the H$\alpha$
emission, but shift the location of the radio emission from the poles
to a hot spot at lower latitude.  We finally note that in any of the
possible configurations, the quiescent non-variable radio emission is
likely produced by the largest scale field structure, such that the
projection effects caused by the rotation and magnetic axis tilt are
minimized.

The possible field topologies that give rise to the observed radio and
H$\alpha$ emission appear to be somewhat more complex than those
inferred for mid-M dwarfs from spectropolarimetric observations
\citep{dfc+06,mdp+08}.  Zeeman-doppler imaging points to predominantly
dipole poloidal fields, particularly for objects with the lowest
Rossby numbers, $Ro\equiv P/\tau_c\lesssim 0.05$, ($\tau_c$ is the
convective overturn timescale).  For \2m0746\ we estimate
$\tau_c\approx (MR^2/L)^{1/3}\approx 200$ d (\S\ref{sec:mr}), so the
inferred Rossby number is extremely small, $Ro\approx 4\times
10^{-4}$, and we may have expected the dipole poloidal configuration
to dominate.  Thus, our observations indicate that the field topology
possibly evolves from mid-M dwarfs to L dwarfs, and that $Ro$ may not
be the only relevant parameter.

\section{The Mass-Radius Relation}
\label{sec:mr}

We next use the rotation period of \2m0746\ as determined from the
radio and H$\alpha$ emission to measure its radius.  We make the
reasonable assumption that the inclination of the rotation axis, $i$,
is identical to the orbital orientation of $142\pm 3^\circ$
\citep{hal94}.  Thus, the projected $v{\rm sin}i=27\pm 3$ km s$^{-1}$
translates to an actual rotation velocity of $v=46\pm 6$ km s$^{-1}$.
For a period of $124.32\pm 0.11$ min, the corresponding radius of
\2m0746\ is $R=(5.4\pm 0.7)\times 10^9$ cm, or $0.078\pm 0.010$
R$_\odot$.  As far as we know, this is the first radius estimate
(without a ${\rm sin}i$ degeneracy) for an L dwarf through photometric
variability.

The total mass of the binary, inferred by \citet{bdk+04} from the
orbital dynamics, is $0.146^{+0.016}_{-0.006}$ M$_\odot$.  These
authors further estimated a range of $0.075-0.095$ M$_\odot$ for the
primary by comparing the infrared photometry with DUSTY model
isochrones.  \citet{gr06}, on the other hand, argued that the primary
mass is $0.078-0.082$ M$_\odot$, primarily as a result of an older age
for the system.  Taking the conservative approach, we adopt here the
wider range of values from \citet{bdk+04}.

\begin{figure}
\centerline{\psfig{file=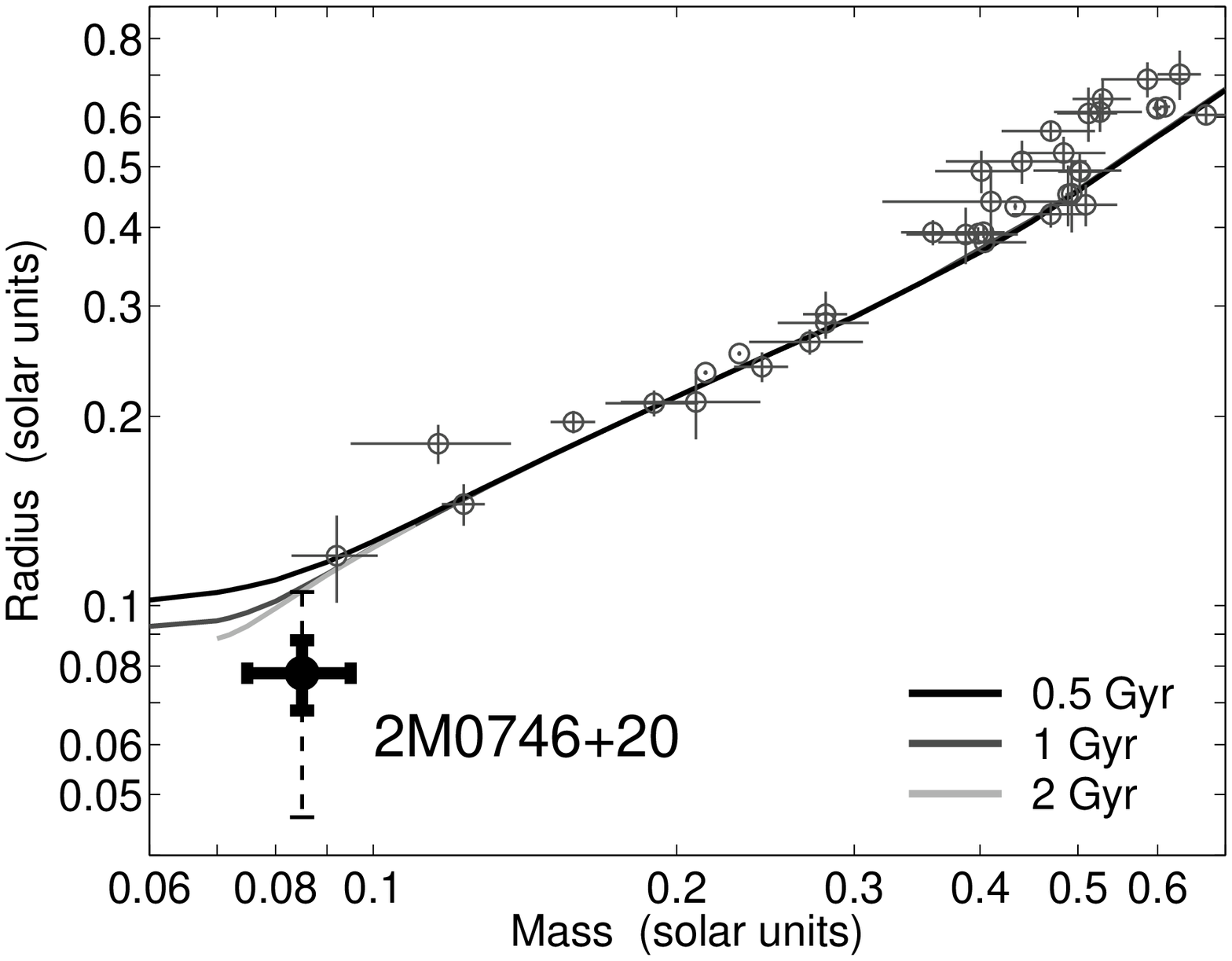,width=6.0in}}
\caption{Stellar radius as a function of mass for the primary star in
\2m0746.  Also shown are model tracks \citep{bca+98} for stellar ages
of 0.5, 1, and 2 Gyr, as well as a compilation of higher mass stars
(open symbols; \citealt{lop07}).  The inferred radius of \2m0746a\ is
about $30\%$ smaller than predicted by the models, with the smallest
discrepancy for the oldest age.  The overall disagreement indicates
either a problem with the evolutionary models near the bottom of the
main sequence, or that our assumption that the rotation and orbital
inclinations are identical is wrong.  The dashed line indicates the
range of inferred radii that corresponds to the minimum allowed radius
(i.e., $i=90^\circ$) and the 90\% inclination probability (i.e., there
is $<10\%$ probability that the inclination angle is such that the
inferred radius will be larger than this value).  This latter value is
marginally consistent with the theoretical models.
\label{fig:mr}} 
\end{figure}

The inferred radius and mass are plotted in Figure~\ref{fig:mr}.  Also
shown are the mass-radius relations from the evolutionary models of
\citet{bca+98} for ages of 0.5, 1, and 2 Gyr and solar metallicity, as
well as masses and radii for stars in the range $0.1-0.7$ M$_\odot$
from a recent compilation by \citet{lop07}.  \2m0746a\ lies below the
model predictions, $R\approx 0.105-0.115$ R$_\odot$, by about $30\%$.
This is surprising since stars with known radii in the mass range
$0.1-0.3$ M$_\odot$ appear to generally agree with the model
predictions \citep{lop07}.  Moreover, stars in the mass range
$0.4-0.7$ M$_\odot$ generally lie above the model predictions,
possibly as a result of magnetic activity \citep{lop07}.  It is
somewhat surprising then, that a highly active L dwarf would underlie
the same theoretical models.  Similarly, a recent estimate of
$R\gtrsim 0.117$ R$_\odot$ for the M8.5 dwarf \lsr\ \citep{had+08},
based on similar periodic radio emission to that of \2m0746, indicates
that there is possibly a larger than expected dispersion in the radii
of ultracool dwarfs.

With only a single object it is impossible to assess whether the
apparent discrepancy indicates that the theoretical models break down
across the sub-stellar boundary, or if this is the result of our
assumption that the rotation axis is aligned with the orbital
inclination .  It is unlikely that the problem is with the value of
$v{\rm sin}i$ since several groups have measured consistent values
(\S\ref{sec:obs}).  If we relax our assumption of $i=142^\circ$, then
the minimum allowed radius (corresponding to $i=90^\circ$) is $0.046$
R$_\odot$.  The radius corresponding to the 90\% probability of the
inclination axis distribution ($i>26^\circ$) is $R\approx 0.105$
R$_\odot$ (i.e., there is only a 10\% probability that $R>0.105$
R$_\odot$).  This latter value is marginally consistent with the
theoretical predictions, but it requires the rotation axis to be
severely misaligned with the orbital inclination.

\section{Summary and Conclusions} 
\label{sec:conc}

We presented simultaneous radio, X-ray, optical, and UV observations
of the L dwarf binary \2m0746.  Both the radio and H$\alpha$ light
curves are dominated by periodic emission, with $P\approx 124$ min.
The agreement between the two periods indicates that both arise from
the same binary member, which we assume to be the primary star.  We
note that high angular resolution Very Long Baseline Array
observations may pinpoint the origin of the radio emission.

The observed period is well-matched to the rotation velocity of
\2m0746, and the assumption that the rotation axis is aligned with the
orbital inclination allows us to infer a radius of $R=0.078\pm 0.010$
R$_\odot$.  Combined with the estimated mass of $0.085\pm 0.010$
M$_\odot$ \citep{bdk+04}, we find that the radius is about 30\%
smaller than predicted from theoretical evolutionary models.  It is
unclear from this single object whether this is the result of a
breakdown in the models, or an indication that the rotation and
orbital axes are severely misaligned.  Both possibilities have
important implications for our understanding of ultracool dwarfs and
their formation mechanism.

The combination of periodic radio and H$\alpha$ emission further
allows us to explore the magnetic field topology.  The radio
periodicity is in the form of short duration pulses with a duty cycle
of $0.8\%$ and a uniform sense of circular polarization, while the
H$\alpha$ emission is sinusoidal and leads the radio emission by 1/4
of a phase.  These emission properties can be explained in the context
of a rotating mis-aligned magnetic field, but the observed phase lag
rules out a symmetric dipole field.  Instead, we conclude that either
the field is dominated by a quadrupole configuration, or the emission
regions are not trivially aligned with a dipole field.  In both
scenarios the quiescent and non-variable radio emission most likely
arises from the largest scales of the field, which are least affected
by rotation and inclination effects.

We note that the availability of simultaneous radio and optical data
is crucial to our understanding of the field geometry.  If only one
band was available to us (as in all previous cases of detected radio
pulses), we would have concluded that the dipole topology is the most
likely scenario.  Instead, the field configuration of \2m0746\ appears
to be somewhat more complex than those inferred for mid-M dwarfs from
phase-resolved optical spectropolarimetry \citep{dfc+06,mdp+08},
although the vast difference in techniques may result in sensitivity
to different field structures.  Regardless of the exact configuration,
our observations support the general conclusions of recent numerical
dynamo simulations \citep{bro08}, which predict large-scale fields at
low Rossby numbers.  Indeed, we estimate that the Rossby number for
\2m0746\ is very low, $\sim 4\times 10^{-4}$.

Future observations of \2m0746\ will allow us to determine the
stability timescale of the magnetic field configuration, a crucial
constraint on dynamo models.  Previous observations of the L3.5 dwarf
2M\,0036+18 \citep{brr+05,had+08} indicated stability on a timescale
of at least $\sim 3$ yr.  In addition, continued surveys for radio
emission from low mass stars and brown dwarfs, particularly in binary
systems, are warranted.  As demonstrated here, these observations,
particularly in conjunction with optical spectroscopy, can provide
important constraints not only on the mass-radius relation below $\sim
0.1$ M$_\odot$, but also on the magnetic field topology of fully
convective objects.

\acknowledgements We thank the Chandra, Gemini, VLA, and Swift
schedulers for their invaluable help in coordinating these
observations.  This work has made use of the SIMBAD database, operated
at CDS, Strasbourg, France.  It is based in part on observations
obtained at the Gemini Observatory, which is operated by the
Association of Universities for Research in Astronomy, Inc., under a
cooperative agreement with the NSF on behalf of the Gemini
partnership: the National Science Foundation (United States), the
Science and Technology Facilities Council (United Kingdom), the
National Research Council (Canada), CONICYT (Chile), the Australian
Research Council (Australia), CNPq (Brazil) and CONICET (Argentina).
Data from the UVOT instrument on Swift were used in this work.  Swift
is an international observatory developed and operated in the US, UK
and Italy, and managed by NASA Goddard Space Flight Center with
operations center at Penn State University.  Support for this work was
provided by the National Aeronautics and Space Administration through
Chandra Award Number G08-9013A issued by the Chandra X-ray Observatory
Center, which is operated by the Smithsonian Astrophysical Observatory
for and on behalf of the National Aeronautics Space Administration
under contract NAS8-03060.


\clearpage

\clearpage

\clearpage

\clearpage

\clearpage

\clearpage

\end{document}